# Modeling the respiratory Central Pattern Generator with resonate-and-fire Izhikevich-Neurons

Pavel Tolmachev [1][0000-0001-6024-0135], Rishi R. Dhingra [2][0000-0002-4684-5215],
Michael Pauley [1][0000-0002-0250-868X], Mathias Dutschmann [2][0000-0002-0692-746X],
and Jonathan H. Manton[1][0000-0002-5628-8563]

[1] Department of Electrical and Electronic Engineering, University of Melbourne, Parkville VIC 3010, Australia
[2] Florey Institute of Neuroscience and Mental Health, 30 Royal Parade, Parkville VIC 3052, Australia
ptolmachev@student.unimelb.edu.au

**Abstract.** Computational models of the respiratory central pattern generator (rCPG) are usually based on biologically-plausible Hodgkin Huxley neuron models. Such models require numerous parameters and thus are prone to overfitting. The HH approach is motivated by the assumption that the biophysical properties of neurons determine the network dynamics. Here, we implement the rCPG using simpler Izhikevich resonate-and-fire neurons. Our rCPG model generates a 3-phase respiratory motor pattern based on established connectivities and can reproduce previous experimental and theoretical observations. Further, we demonstrate the flexibility of the model by testing whether intrinsic bursting properties are necessary for rhythmogenesis. Our simulations demonstrate that replacing predicted mandatory bursting properties of pre-inspiratory neurons with spike adapting properties yields a model that generates comparable respiratory activity patterns. The latter supports our view that the importance of the exact modeling parameters of specific respiratory neurons is overestimated.

## 1 Introduction

Respiration is one of the vital processes of life. While in simple single cell organisms respiration is driven by passive diffusion, complex organisms have developed complex breathing organs for the uptake of atmospheric oxygen and excretion of $CO_2$ (e.g. gills and lungs). In mammals, airflow in and out of the lungs is generated by a variety of respiratory thoracic and abdominal muscles [1], while the strength and duration of pulmonary airflow is regulated by valving muscles in the upper airways [2, 3]. The former include the diaphragm, the primary inspiratory muscle, expiratory and intercostal and finally expiratory abdominal muscles. The latter include laryngeal adductor and abductor muscles, as well as the tongue and various muscles of the soft palate and pharynx.

Besides the bronchomotor muscles, all respiratory muscles are skeletal and are therefore controlled by the brain. The brain breathing centers of mammals are organized in neuronal columns that span the medulla oblongata and the pons, which form the anatomical substrate for the central respiratory pattern generator (rCPG). Specific compartments of the rCPG are seen to serve a specific function in respiratory rhythm generation and formation of a three-phase sequential motor pattern compromising inspiration, postinspiration (stage I expiration) and expiration (stage II expiration) [4-11].

Over the last century, experimental data accumulated that identified the basic behavior of respiratory neurons that are distributed within specific compartments of rCPG. The class of neurons traditionally encodes the phase of neuronal activity in relation to the inspiratory activity of the diaphragm or the phrenic nerve. In addition, augmenting and decrementing discharge frequencies of these neurons are considered for classification. There is a general consensus that 5 classic respiratory neuron types form the core of the neural circuit that generates the respiratory rhythm and motor pattern: (1) rhythmogenic pre-Inspiratory (pre-I), (2) early-Inspiratory (early-I) with a decrementing discharge pattern (thus also called I-Dec), (3) Inspiratory neurons with augmenting discharge pattern (I-Aug, or ramp-I), post-Inspiratory neurons (post-I) with decrementing discharge pattern, which are active during the first part of expiration (thus also called E-Dec) and finally expiratory neurons that show augmenting discharge (E-Aug) pattern during the second phase of the expiratory interval. These neuron types form the basis for a substantial number of computational models that describe the putative function of the r-CPG. Mathematical models have focused on several dynamical mechanisms including the biophysical bursting properties of rhythmogenic pre-I that are seen to initiate the respiratory cycle [12-14], network oscillation based on reciprocal synaptic inhibition [15] and hybrid models based on excitatory rhythmogenic cell properties and inhibitory synaptic inhibition [7, 16, 23]. The latter sometimes even implement sensory feedback loops [17, 18]. The contemporary hybrid models are complex Hodgkin-Huxley-based models. The main driver for Hodgkin-Huxley-based modeling has largely arisen from the finding in the early nineties that a specific subset of neurons located in the pre-Bötzinger complex (pre-BötC) remain rhythmogenic when isolated from the larger network [19]. Electrophysiological studies of pre-BötC neurons revealed the biophysical basis of pacemaker pre-I neurons and the excitatory synaptic coupling underlying group pacemaker mechanisms [20]. However, the biophysical properties of neurons outside the pre-BötC remain largely unexplored. Thus, the biophysical properties of non-pre-BötC neurons in modeling approaches are often based on speculation. Even more compelling is that fact that biophysical properties (ion channel composition) of bursting neurons in the somatogastric ganglion, an invertebrate model system of a CPG network, are extremely diverse in functionally homogenous neurons and therefore do not define the function of neurons in a rhythmogenic circuit [21] that shares significant similarity with the rCPG.

To simplify the complex and difficult to assess Hodgkin-Huxley-based models of the respiratory CPG, we implement the rCPG using Izhikevich resonate-and-fire neurons. These neurons are simple enough to be computationally efficient and tractable for bifurcation analysis and at the same time can account for various neural activity patterns observed within the respiratory circuit *in vivo*, including intrinsic bursting and spike adaptation [22].

The remainder of the manuscript is organized as follows. In Section 2, we present a description of the model. In Section 3, we present our results regarding how the model can reproduce previous experimental and theoretical observations, and how the model can be used to test the hypothesis that intrinsically bursting neurons are necessary for the expression of the three-phase respiratory rhythm. Finally, in Section 4, we present our overall conclusions and discuss the utility of the model for future research on respiratory rhythm and pattern generation.

## 2  Modeling description

### 2.1  Izhikevich neurons

To construct a model of the rCPG, which is flexible and more transparent than models utilizing Hodgkin-Huxley neurons, we employed Izhikevich neurons. The model of an Izhikevich neuron consists of two differential equations, resulting in a 2-dimensional system. The phase plane trajectory of these neurons is thereby easily visualized, making it tractable for bifurcation analysis. This simplifies the task of finding parameters to achieve a desired neuronal behavior.

The general form of the equations is presented below:

$$\frac{dv}{dt} = \alpha(v - v_0)^2 + V_b - xu \tag{1}$$

$$\frac{du}{dt} = a(bv - u) \tag{2}$$

with two additional resetting conditions as the spike occurs:

$$if\ V = V_{threshold} : V \to V_{reset}; u \to u + d \tag{3}$$

Here $v$ is the membrane potential in mV, and $u$ is an adaptation variable. The greater the parameter $u$, the slower the voltage rises after resetting. $\alpha$, $v_0$, $V_b$, $a$, $b$, $x$ are the parameters of the system whose geometric meaning can be seen in **Fig. 1**. The sensitivity of the membrane potential to adaptation is governed by parameter $x$. Parameter $\alpha$ is proportional to the width of the parabolic-shaped $v$-nullcline. $v_0$ and $V_b$ define the coordinates of the peak of the $v$-nullcline. The slope of $u$-nullcline is set by a parameter $b$. The parameter $a$ is responsible for the rate of relaxation of the adaptation variable.

Our model of the rCPG includes only two sets of parameters for a single neuron to reproduce the two different neural behaviors critical to the established connectivity of the rCPG: spike adaptation—a gradual decline of the frequency given constant tonic input—and intrinsic bursting—the alternation of spiking and quiescent periods due to an oscillation in the slow-subsystem. By visualizing the phase plane of Izhikevich neurons, we were able to find suitable sets of parameters to account for these properties (see Appendix). We also ensured that these parameter sets were associated with biologically plausible firing frequencies.

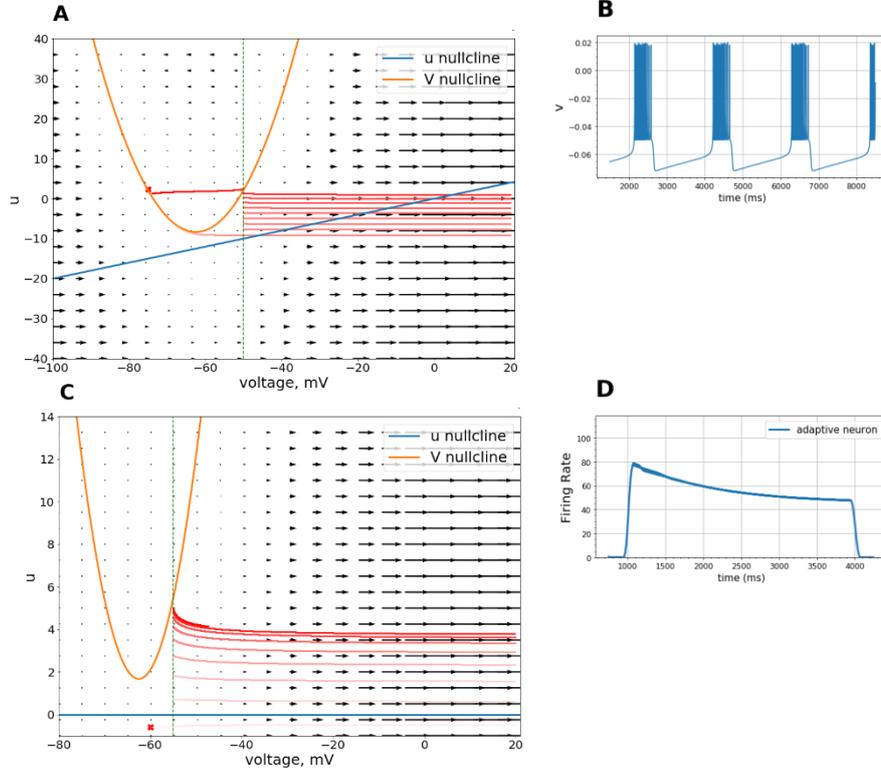

**Fig. 1.** Examples of a phase plane diagram and the corresponding behaviour of the Izhikevich model for a bursting (top: **A,** with *v* plotted against the time in **B**) and adaptively spiking neuron (bottom: **C,** with firing rate plotted against time in **D**). *V*-nullcline from the expression (1) has a parabolic shape. The *u*-nullcline is a straight line. The arrows depict the vector field of $(\frac{dv}{dt}, \frac{du}{dt})$, where derivatives are the right-hand side of the expressions (1) and (2). The green dashed line indicates the resetting voltage in equation (3). The red line indicates the phase trajectory of the neuron: the brighter the line the more recently the system was in that point of the phase-plane. The cross marks indicate the initial conditions that the system evolved from. The adaptation of firing rate (C) occurs because initially a neuron's state is in the region where the time derivative of voltage is larger. The decline of the firing rate continues until the neuron reaches a dynamic equilibrium: after each spike the adaptation parameter is incremented by value ***d***, but during the spike initiation it decays back by the same amount. The derivative of voltage along the resulting stationary trajectory is less than in the regions with smaller ***u***, resulting in the decreased firing rate. In case of bursting (A), the dynamical equilibrium is not reached: eventually, after some spike, the system's state goes into a region inside the parabola. That moment in time marks the end of bursting. The state of the neuron evolves in the phase-space according to a vector field, and the neuron's voltage drops to the minimum when the state reaches the left branch of the parabola. The neuron initiates a new bursting after the adaptation parameter relaxes below the parabola's peak. Our rCPG model is composed strictly of neurons with one of these behaviors.

To account for the synaptic interactions between the neurons, one has to introduce an extra term into equation (1):

$$I_{syn} = g(v - E_{syn}) \quad (4)$$

where $g$ is the conductance of the synapses of a particular neuron and is comprised of two parts:

$$g = g_{net} + g_{tonic} \quad (5)$$

The parameter $g_{net}$ is the network contribution into the overall conductance of the synapses, $g_{tonic}$ is the baseline conductance due to activity of the tonic drive populations. $E_{syn}$ is the synaptic reversal potential: for excitatory synapses, $E_{syn}$ is around *-10 mV*, for inhibitory synapses, $E_{syn}$ is approximately *-70 mV*. The synaptic conductance $g_{net}$, in turn, is subject to the following differential equation with an extra condition on the arrival of the spike from the presynaptic neuron:

$$\frac{dg_{net}}{dt} = \frac{-g_{net}}{\tau} \quad (6)$$

$$if\ spike \in presynaptic\ neuron: g_{net} \rightarrow g_{net} + \Delta \quad (7)$$

Thus, the network contribution to the synaptic conductance exponentially decays back to 0, unless the spike arrives from an afferent neuron. The arrival of the spike causes the conductance to swiftly rise by a level defined by parameter Δ (which represents the synaptic efficacy).

All simulations were implemented using the *Brian2* simulation package.

## 2.2 Model description

The baseline synaptic connectivity in our model implemented the conceptual model described in [1].

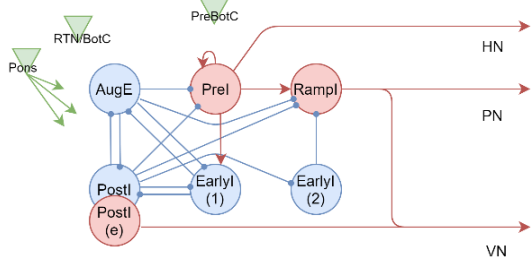

**Fig. 2.** Network connectivity between six populations of neurons of the rCPG. Blue circles denote inhibitory populations, whereas red circles represent excitatory populations. Post-I neurons have both inhibitory and excitatory subpopulations, which receive almost identical inputs from the rest of the network. Inhibitory connections are depicted by blue lines with a filled circle. The red arrows signify excitation between neural groups. All of the groups receive tonic input from three major sources: PreBötC, RTN/BötC and the pontine compartments. These tonic drives modify the conductance of the neurons by setting a baseline conductance $g_{tonic}$. The parameters for the tonic drives for each neuronal group are described in the Table 2 (see Appendix). The activities of respiratory neurons are projected to three nerve fibres: Hypoglossal nerve (HN), Phrenic nerve (PN) and Vagal nerve (VN).

The effects of synaptic plasticity are considered to be negligible in CPGs, and in our models, the synaptic strengths for all synapses were set equal. The probabilities of connection between the neuronal populations are listed in Table 1 (see Appendix). The tonic drives for the various groups of neurons are shown in Table 2. In our simulations, each neural group represents a population of 100 neurons with identical parameters, which are described in Table 3 (see Appendix). In the simulations, we have set random initial conditions for the neurons in the model. To account for the heterogeneity, we have also introduced some variability of the synapses' responses to spikes. The parameters $d$, an increment of the parameter $u$ after the arrival of a presynaptic spike, and $\Delta$, or the synaptic efficacy, are normally distributed random variables with the variance equal 10% of the mean.

In the model, we set the activity of HN identical to the activity of pre-I neurons. The PN activity is set to equal to the activity of ramp-I neurons. The activity of VN was described as the combination of activities of post-I(e) neurons in conjunction with ramp-I neurons with the weighted coefficients 0.75 and 0.25, respectively.

## 3    Results

### 3.1    rCPG replicates the properties of Hodgkin-Huxley-based models

Under normal metabolic conditions, the circuit operates in the three-phase mode described in [23]. Our model demonstrates results consistent with both experimental observations and the model based on the Hodgkin-Huxley neurons [23]. The results of our model simulations are presented in Fig. 3 and are summarized in the next paragraph.

Post-I neurons escape from inhibition by early-I (1) neurons and start to fire. Their firing rate declines through the $E_1$ phase and reaches the point when the post-I activity is no longer sufficient to fully inhibit aug-E neurons. Aug-E neurons exhibit an incrementing discharge pattern during the $E_2$ stage, while the activity of the post-I neurons continues to decline. Eventually, inhibition of pre-I neurons by aug-E and post-I neural groups can no longer be maintained, and pre-I neurons start to fire. Since pre-I neurons excite the early-I neurons, the latter also escape from inhibition arising from aug-E and post-I neurons. Early-I neurons then inhibit aug-E and post-I groups and fire throughout the inspiratory phase. An intrinsic adaptation of early-I neurons results in decrementing firing rate until early-I activity is no longer sufficient to inhibit the post-I population. Then, the cycle repeats.

Our model of the rCPG, although lacking its biophysical properties, robustly reproduces the modeling results of similar Hodgkin-Huxley-based models. The discharge patterns of the neurons are also consistent with experimental data, except for the shape of activity of the pre-I neurons, for which the experimental data suggests that the firing rate of pre-I neurons is incrementing. A possible solution to reconcile this discrepancy is also purely mechanistic and not dependent on a neuron's biophysics: one may introduce into the model a new group of decrementing inhibitory neurons, which fire in the inspiratory phase and inhibit the pre-I neurons (data not shown).

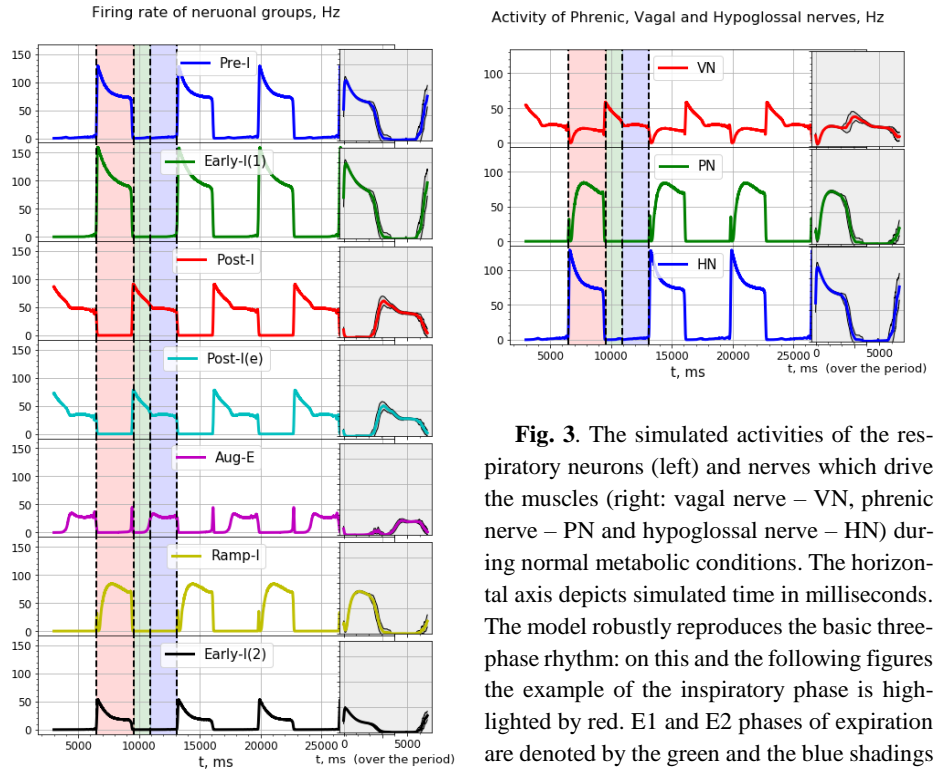

**Fig. 3**. The simulated activities of the respiratory neurons (left) and nerves which drive the muscles (right: vagal nerve – VN, phrenic nerve – PN and hypoglossal nerve – HN) during normal metabolic conditions. The horizontal axis depicts simulated time in milliseconds. The model robustly reproduces the basic three-phase rhythm: on this and the following figures the example of the inspiratory phase is highlighted by red. E1 and E2 phases of expiration are denoted by the green and the blue shadings respectively. The gray shaded insets on the right represent the firing rate averaged consecutively (1) over the period and (2) over 50 different trials. The variability of a signal within one standard deviation is depicted on each inset by a shading between two black lines with mean laying in-between. The variability of the signal comes from 1) randomness in initial conditions and 2) stochastic character of the connectivity. The average period T = (6372 ± 675) ms, CV = 0.106. The firing patterns of the neurons are consistent with those described in [23].

### 3.2 Simulations of pontine transection are consistent with experimental data and previous models

To further test the validity of our model, we have simulated pontine transection, which experimentally causes apneusis—a breathing pattern characterized by prolonged inspiration and an absence of the post-inspiratory phase activity. A representative example of this simulation is presented in Fig. 4.

Because post-I neuronal activity greatly depends on pontine drive, removal of pontine drive to post-I neurons causes post-I neurons to become quiescent. Additionally, Aug-E neurons are released from the inhibition from the post-I population, and, in conjunction with early-I neurons, these two groups form a half-center oscillator. The oscillator produces a two-phase apneustic respiratory rhythm. Our simplified model produces neuronal and motor outputs which have the same patterns as earlier models and is also consistent with the data obtained from experiments.

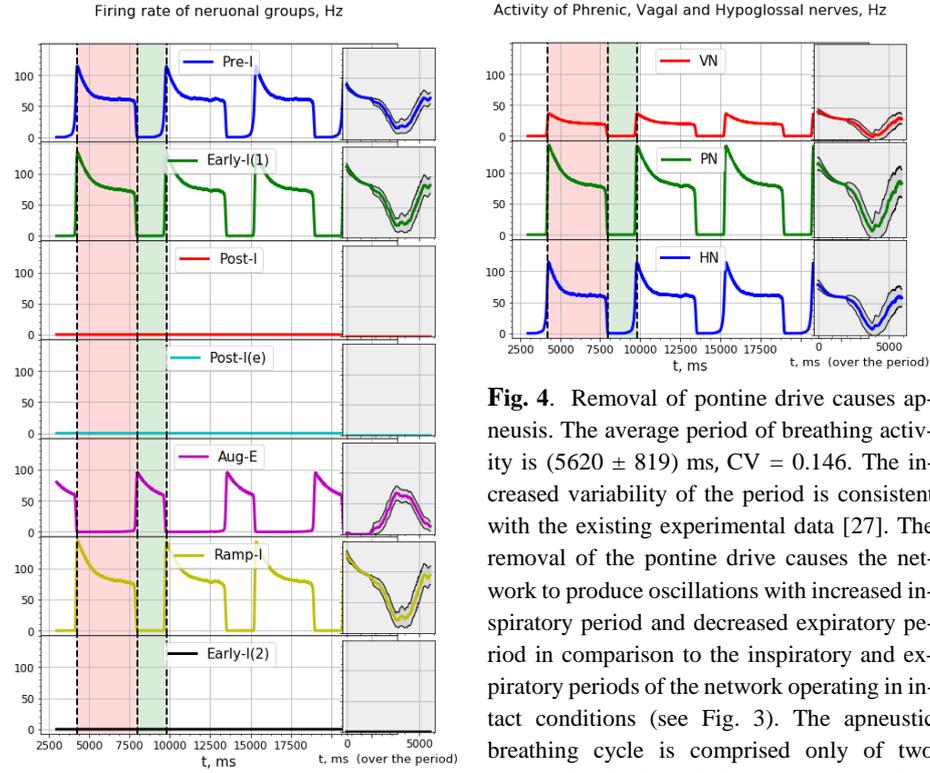

**Fig. 4**. Removal of pontine drive causes apneusis. The average period of breathing activity is (5620 ± 819) ms, CV = 0.146. The increased variability of the period is consistent with the existing experimental data [27]. The removal of the pontine drive causes the network to produce oscillations with increased inspiratory period and decreased expiratory period in comparison to the inspiratory and expiratory periods of the network operating in intact conditions (see Fig. 3). The apneustic breathing cycle is comprised only of two phases: inspiration (depicted by red shading) and expiration (denoted by green shading). The insets with grey shading depict one period of the activity and show mean and the standard deviation of simulations averaged over 50 trials. This apneustic breathing pattern is consistent with the experimental data.

### 3.3 Intrinsically bursting properties are not essential to generate the three-phase respiratory rhythm

Using the simplified model, we were able to test whether intrinsic bursting properties were necessary for the generation of the three-phase rhythm by introducing only minor changes to the model. In this section, we discuss the effect of switching pre-I neurons' dynamics from a pacemaker to an adaptively spiking parameter set (Fig. 5). To adjust the period of inspiration we have also decreased tonic drive to pre-I neurons. Our simulations show that the network may produce the three-phased rhythm even with non-bursting pre-I neurons. According to our model of the respiratory CPG, switching the dynamics of pre-I neurons did not result in qualitative changes. Even with adaptively spiking pre-I neurons, the network still produced a the three-phase rhythm similar to that produced by the unperturbed model and described in the section 3.1. Our results therefore suggest that the role of the pacemaker neurons in the network under normal metabolic conditions may be overestimated.

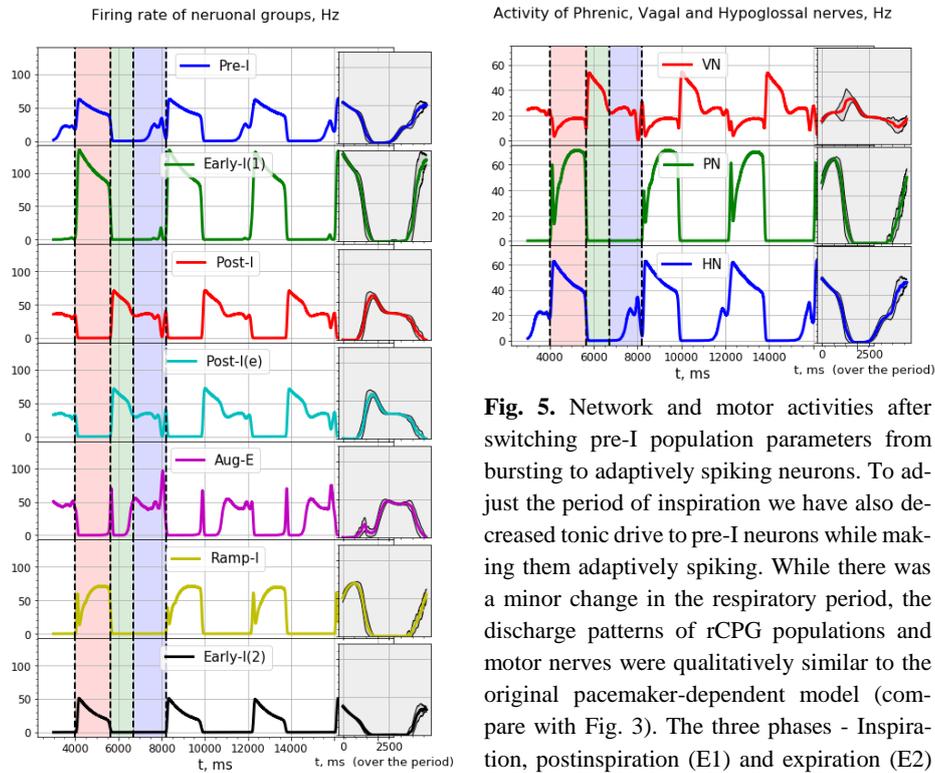

**Fig. 5.** Network and motor activities after switching pre-I population parameters from bursting to adaptively spiking neurons. To adjust the period of inspiration we have also decreased tonic drive to pre-I neurons while making them adaptively spiking. While there was a minor change in the respiratory period, the discharge patterns of rCPG populations and motor nerves were qualitatively similar to the original pacemaker-dependent model (compare with Fig. 3). The three phases - Inspiration, postinspiration (E1) and expiration (E2) are denoted by red, green and blue shadings respectively. The average period of a respiratory cycle T = (4380 ± 227) ms, CV = 0.052.

## 4      Conclusion

We conclude that (1) the simplified model reproduces the three-phase respiratory pattern, (2) the model reproduces the effect of the removal of pontine components of the rCPG, and (3) intrinsic bursting properties are not necessary to generate the three-phase respiratory pattern given the currently accepted connectivity of the rCPG.

The respiratory cycle consists of three sequential 'phases': inspiration, post-inspiration and expiration. Neuronal populations within the rCPG are active during specific respiratory phases. Like previous models [7, 17, 23], our simplified resonate-and-fire model largely reproduces this distinct respiratory motor pattern. Employing the resonate-and-fire neurons allowed us to focus on connectivity, rather than biophysics.

However, the current model also shows some instability during the phase transitions that are reflected by short bursts that are quickly terminated when the phase irreversibly changes (see Fig. 3). This is likely due to a competition between the three reciprocally-connected inhibitory populations that arises from the variability of synaptic connection probabilities (see Table 1). For a short period of time during the switching, combined activitiy of pre-I and post-I groups is not enough to inhibit the aug-E population. This feature was also seen in similar computational models [26]. In the future, we expect to explore the parameter space of possible connectivities using a genetic search algorithm and will remove this property from the model to simplify the analysis. On the other hand, similar transition instabilities can be observed experimentally when pontine components of the network are absent. Previous work suggested that such aberrant phrenic nerve discharges may be due to the widespread excitatory projections of medullary pre-I neurons that trigger synchronous bursting in functionally different respiratory motor output [24].

The current model reproduces the effect of pontine transection or pontine synaptic blockade as it has been observed experimentally [8, 23, 25] and in the corresponding models [7, 17, 23]. In accordance with previous models, this was achieved by removing pontine tonic synaptic input to the early-I and post-I populations. Not surprisingly, this eliminates the synaptic instability during phase transitions. In the future, we hope to use this model to explicitly consider alternative connectivities between medullary and pontine respiratory neurons, which are currently oversimplified.

Interestingly, we observed that replacing intrinsically bursting pre-I neurons with adaptively spiking pre-I neurons did not qualitatively change the three-phase rhythm. This supports our view that biophysical properties of a specific subset of respiratory neurons might be overstated. Thus, we aim to utilize the current model for further investigation of the dependence of respiratory rhythm on respiratory circuit connectivity. While part of our motivation for the current model was to reduce the parameter space of the model, because our model includes many neurons, it ultimately is still prone to overfitting. Others have accomplished this task more effectively [26] by developing a squashing function to reduce a population of identical Hodgkin-Huxley neurons with slow synaptic transmission into a single Hodgkin-Huxley neuron's firing rate. This allowed for bifurcation analysis at a network level to understand the dynamic regulation of phase transition, the hallmark mechanism for respiratory pattern formation that defines breathing.

## 5    Appendix

**Table 1.** Synaptic connectivities.

| Synapse | Probability |
|---|---|
| Pre-I → Pre-I | 0.125 |
| Pre-I → Early-I(1) | 0.8 |
| Aug-E → Pre-I | 0.06 |
| Aug-E → Early-I(1) | 0.5 |
| Early-I(1) → Aug-E | 0.5 |
| Aug-E → Early-I(1) | 0.5 |
| Post-I → Early-I(1) | 0.5 |
| Post-I → Aug-E | 0.7 |
| Early-I(1) → Post-I | 0.5 |
| Aug-E → Post-I | 0.1 |
| Post-I → Pre-I | 0.15 |
| Aug-E → Post-I(e) | 0.13 |
| Early-I(1) → Post-I(e) | 0.5 |
| Pre-I → Ramp-I | 0.625 |
| Early-I(1) → Ramp-I | 0.625 |
| Aug-E → Ramp-I | 0.5 |
| Post-I → Ramp-I | 0.2 |
| Early-I(2) → Ramp-I | 0.8 |
| Aug-E → Early-I(2) | 0.2 |

**Table 2.** Tonic drives to neuronal groups.

| Drive → Population ↓ | from PreBötC | from RTN/BötC | from Pons |
|---|---|---|---|
| Pre-I | 0.1 | 0.2 | 0.3 |
| Early-I(1) | 0 | 0.6 | 0.5 |
| Aug-E | 0 | 1 | 0.8 |
| Post-I | 0 | 0 | 0.9 |
| Post-I(e) | 0 | 0 | 0.6 |
| Ramp-I | 0 | 0 | 0 |
| Early-I(2) | 0 | 0 | 0.2 |

**Table 3.** Parameters for a single neuron.

| Population | Bursting | Adaptation |
|---|---|---|
| $\alpha$ | 0.004 | 0.004 |
| $v_0$ | -62.5 | -62.5 |
| $V_b$ | -1.6 | 0.0 |
| $a$ | 0.001 | 0.0005 |
| $b$ | 0.2 | 0.0 |
| $E_{synE}$ | -10 | -10 |
| $\tau_E$ | 10 | 10 |
| $g_{netE}$ | 0.1 | 0.33 |
| $g_{tonicE}$ | 0.1 | 0.1 |
| $E_{synI}$ | -75 | -75 |
| $\tau_I$ | 15 | 15 |
| $g_{netI}$ | 0.1 | 1.0 |
| $\Delta$ | 0.08 | 0.08 |
| $V_{reset}$ | -50 | -55 |
| $V_{threshold}$ | 20 | 20 |
| $d$ | 0.3 | 0.5 |
| $x$ | 0.06 | 0.06 |

## References

1. Feldman, J.L.: Neurophysiology of breathing in mammals. Handbook of physiology. The nervous system. Am. Physiol. Soc., Sect, 1, 463-524 (1986)
2. Dutschmann, M., Paton J.F.: Inhibitory synaptic mechanisms regulating upper airway patency. Respiratory physiology & neurobiology 131.1-2: 57-63 (2002)
3. Dutschmann, M., Jones, S.E., Subramanian, H.H., Stanic, D., & Bautista, T.G.: The physiological significance of postinspiration in respiratory control. Prog. Brain Res. 212,113-130 Elsevier (2014)
4. Richter, D.W.: Generation and maintenance of the respiratory rhythm. J. Exp. Biol., 100(1), 93-107 (1982)
5. Richter, D.W., & Spyer, K. M.: Studying rhythmogenesis of breathing: comparison of in vivo and in vitro models. Trends Neurosci., 24(8), 464-472 (2001)
6. Feldman, J.L., & Del Negro, C.A.: Looking for inspiration: new perspectives on respiratory rhythm. Nature Rev Neurosci, 7(3), 232 (2006)
7. Rybak, I.A., Abdala, A.P., Markin, S.N., Paton, J.F., & Smith, J.C.: Spatial organization and state-dependent mechanisms for respiratory rhythm and pattern generation. Prog. Brain Res., 165, 201-220 (2007)
8. Dutschmann, M., & Dick, T.E.: Pontine mechanisms of respiratory control. Compr. Physiol., 2(4), 2443 (2012).

9. Smith, J.C., Abdala, A.P., Borgmann, A., Rybak, I.A., & Paton, J.F.: Brainstem respiratory networks: building blocks and microcircuits. Trends Neurosci., 36(3), 152-162 (2013)
10. Anderson, T.M., & Ramirez, J.M.: Respiratory rhythm generation: triple oscillator hypothesis. F1000Research, 6 (2017)
11. Del Negro, C.A., Funk, G.D., & Feldman, J.L.: Breathing matters. Nat Rev Neurosci. (2018)
12. Butera Jr, R.J., Rinzel, J., & Smith, J.C.: Models of respiratory rhythm generation in the pre-Botzinger complex. I. Bursting pacemaker neurons. J. Neurophysiol., 82(1), 382-397 (1999)
13. Butera Jr, R.J., Rinzel, J., & Smith, J.C.: Models of respiratory rhythm generation in the pre-Botzinger complex. II. Populations of coupled pacemaker neurons J. Neurophysiol., 82(1), 398-415 (1999)
14. Del Negro, C.A., Johnson, S.M., Butera, R.J., & Smith, J.C.: Models of respiratory rhythm generation in the pre-Botzinger complex. III. Experimental tests of model predictions. J. Neurophysiol., 86(1), 59-74 (2001)
15. Ogilvie, M.D., Gottschalk, A., Anders, K., Richter, D.W., & Pack, A.I.: A network model of respiratory rhythmogenesis. Am. J. Physiol. Regul. Integr. Comp. Physiol., 263(4), R962-R975 (1992)
16. Smith, J.C., Abdala, A.P.L., Koizumi, H., Rybak, I.A., & Paton, J.F.: Spatial and functional architecture of the mammalian brain stem respiratory network: a hierarchy of three oscillatory mechanisms. J. Neurophysiol., 98(6), 3370-3387 (2007)
17. Rybak, I.A., Shevtsova, N.A., Paton, J.F.R., Dick, T.E., John, W.S., Mörschel, M., & Dutschmann, M.: Modeling the ponto-medullary respiratory network. Respir. Physiol. Neurobiol., 143(2-3), 307-319 (2004)
18. Molkov, Y.I., Bacak, B.J., Dick, T.E., & Rybak, I.A.: Control of breathing by interacting pontine and pulmonary feedback loops. Front. Neural Circuits, 7, 16 (2013)
19. Smith, J.C., Ellenberger, H.H., Ballanyi, K., Richter, D.W., & Feldman, J.L.: Pre-Botzinger complex: a brainstem region that may generate respiratory rhythm in mammals. Science, 254(5032), 726-729 (1991)
20. Del Negro, C.A., Morgado-Valle, C., & Feldman, J.L.: Respiratory rhythm: an emergent network property? Neuron, 34(5), 821-830 (2002)
21. Schulz, D.J., Goaillard, J.M., & Marder, E.: Variable channel expression in identified single and electrically coupled neurons in different animals. Nat. Neurosci., 9(3), 356 (2006)
22. Izhikevich, E.M.: Simple model of spiking neurons. IEEE Trans. Neural Netw., 14(6), 1569-1572 (2003)
23. Smith, J.C., Abdala, A.P.L., Koizumi, H., Rybak, I.A., & Paton, J.F.: Spatial and functional architecture of the mammalian brain stem respiratory network: a hierarchy of three oscillatory mechanisms. J. Neurophysiol., 98(6), 3370-3387 (2007)
24. Jones, S.E., & Dutschmann, M.: Testing the hypothesis of neurodegeneracy in respiratory network function with a priori transected arterially perfused brain stem preparation of rat. J. Neurophysiol., 115(5), 2593-2607 (2016)
25. Dhingra, R.R., Jacono, F.J., Fishman, M., Loparo, K.A., Rybak, I.A., & Dick, T.E.: Vagal-dependent nonlinear variability in the respiratory pattern of anesthetized, spontaneously breathing rats. J. Appl. Physiol., 111(1), 272-284 (2011)
26. Rubin, J.E., Shevtsova, N.A., Ermentrout, G.B., Smith, J.C., & Rybak, I.A.: Multiple rhythmic states in a model of the respiratory central pattern generator. J. Neurophysiol., 101(4), 2146-2165 (2009)
27. Dhingra, R.R., Dutschmann, M., Galán, R.F., & Dick, T.E.: Kölliker-Fuse nuclei regulate respiratory rhythm variability via a gain-control mechanism. Am. J. Physiol. Regul. Integr. Comp. Physiol., 312(2), R172-R188 (2016)